\documentclass[a4paper]{jpconf}
\usepackage{graphicx}
\usepackage{gensymb}
 \usepackage{color}
 \usepackage{hyperref} 
\usepackage{amsmath}
\usepackage{amssymb}
\usepackage{pifont}
\usepackage{latexsym}
\usepackage{dsfont }
\usepackage{float}
\usepackage{multirow}
\usepackage{verbatim}
\usepackage{tikz}

\def\bk{{\bold{k}}}
\def\bh{{\bold{h}}}

\def\up{{\uparrow}}
\def\dn{{\downarrow}}

\def\m1{{^{-1}}}

\begin{document}
\title{Nodal gaps from local interactions in Sr$_2$RuO$_4$}

\author{Aline Ramires}

\address{Paul Scherrer Institut, CH-5232 Villigen PSI, Switzerland}

\ead{aline.ramires@psi.ch}

\begin{abstract}
Sr$_2$RuO$_4$ has been under intensive scrutiny over the past years after new NMR measurements unveiled that the superconducting state might be a spin singlet. One of the best order parameter candidates in light of these new experiments is a chiral d-wave state with $d_{xz}+id_{yz}$ symmetry. This order parameter has been overlooked given the strong two-dimensional character of the normal state electronic structure 
Recently, a phenomenological proposal based on local interactions with a three-dimensional electronic structure showed that a chiral d-wave state can be stable in Sr$_2$RuO$_4$ once momentum-dependent spin-orbit coupling is properly taken into account. Here we discuss the origin of the nodes and dips in this order parameter as inherited from the normal state Hamiltonian, showing that a nodal gap can emerge out of purely local interactions and connect the presence of nodes with the superconducting fitness measure.
\end{abstract}


\section{Introduction}

Sr$_2$RuO$_4$ was for long thought to be the best candidate material to host a chiral p-wave superconducting state \cite{Rice:1995, Mackenzie:2017}. Recent Knight shift experiments, displaying a significant reduction of the spin susceptibility across the superconducting transition usually associated with spin singlet states, challenged this proposal  \cite{Pustogow:2019,Ishida:2020}. Complementary experiments such as Kerr effect \cite{Xia:2006} and muon spin resonance \cite{Luke:1998,Grinenko:2021} indicate time-reversal symmetry breaking in the superconducting state, and ultrasound attenuation experiments \cite{Lupien:2001,Benhabib:2021,Ghosh:2021} suggest a multi-component order parameter. Together, these experiments suggest an order parameter in the $E_g$ irreducible representation of the $D_{4h}$ point group. This order parameter has two components, one with $d_{xz}$ and another with $d_{yz}$ symmetry. In its chiral form, usually referred to as chiral d-wave, or $d_{xz}+id_{yz}$ state, the order parameter breaks time-reversal symmetry. This order parameter candidate was for a long time overlooked within the assumption that a superconducting state with horizontal line nodes requires strong inter-layer interactions, what is at a first sight in contradiction with the strongly two-dimensional Fermi surfaces of Sr$_2$RuO$_4$ \cite{Bergemann:2003}.

Recently, the author and collaborators have proposed a picture with local pairing in which a chiral d-wave order parameter is stable \cite{Suh:2020}. The main ingredients in this phenomenological theory are strong Hund's coupling and a faithful description of the normal state in three dimensions, including momentum dependent spin-orbit coupling. The projected gap at the Fermi surfaces have horizontal line nodes at the $k_z=\{0,\pm 2\pi\}$ planes, and minima along the diagonals $k_x = \pm k_y$, reminiscent of a $d_{x^2-y^2}$ gap. The aim of this proceeding article is to discuss how these qualitative features of the gap come about. The main aspect to be highlighted is the construction of the normal state and superconducting order parameters from a microscopic perspective, within the orbital basis. Once the analysis is translated to the band basis, these unexpected features emerge.



\section{The normal state Hamiltonian}\label{Sec:H0}

Following the notation introduced in \cite{Suh:2020}, we can describe the normal state of Sr$_2$RuO$_4$ by a three-dimensional three-orbital model. We define the spinor operator $\Phi^\dagger_{\bk} = (c_{\bk,yz\up}^\dagger,\allowbreak c_{\bk,yz\dn}^\dagger,\allowbreak c_{\bk,xz\up}^\dagger,\allowbreak c_{\bk,xz\dn}^\dagger,\allowbreak c_{\bk,xy\up}^\dagger,\allowbreak c_{\bk,xy\dn}^\dagger)$, where $c^\dagger_{\bk,\gamma s}$ creates an electron with momentum $\bk$, orbital $\gamma$, and spin $s$, in terms of which we can write the most general single-particle Hamiltonian
$
\mathcal{H}_0=\sum_{\bk}\Phi^\dagger_\bk \hat{H}_0(\bk) \Phi_\bk
$, 
with
$
\hat{H}_0(\bk) = \sum_{a=0}^8\sum_{b=0}^3 h_{ab}(\bk)\, \hat{\lambda}_a \otimes \hat{\sigma}_b$. 
Here $\hat{\lambda}_a$ are Gell-Mann matrices encoding the orbital degrees of freedom (DOF), $\hat{\sigma}_b$ are Pauli matrices encoding the spin DOF
 ($\hat{\lambda}_0$ and $\hat{\sigma}_0$ are unit matrices), and $h_{ab}(\bk)$
are real and even functions of  momentum. In presence of spacial inversion and time-reversal symmetries, out of the 36 pairs of indexes $(a,b)$, only 15 are symmetry allowed. These are described in detail in the supplementary material of Suh et al. \cite{Suh:2020}. With parameters obtained from first principle calculations, this model generates three almost cylindrical Fermi surfaces with a non-trivial orbital and spin texture (see for example  Veenstra et al. \cite{Veenstra:2014}).

For the analytical discussion below, it is useful to focus on the normal state Hamiltonian along specific planes. If we look along the $XZ$ ($YZ$) plane we find only the $\beta$ and $\gamma$ bands, the former composed of $d_{xz}$ ($d_{yz}$) and the later of $d_{xy}$ orbitals. This allows us to construct simpler two-orbital models along these planes,
$
\mathcal{H}_0^{XZ(YZ)}=\sum_{\bk}\Psi^\dagger_\bk \hat{H}_0^{XZ(YZ)}(\bk) \Psi_\bk,
$
in the basis $\Psi^\dagger_\bk = (c_{\bk,xz(yz)\up}^\dagger,\allowbreak c_{\bk,xz(yz)\dn}^\dagger,\allowbreak
c_{\bk,xy\up}^\dagger,\allowbreak c_{\bk,xy\dn}^\dagger)$, with
$
\hat{H}_{0}^{XZ(YZ)}(\bk) = \sum_{a,b =0}^3 \tilde{h}_{ab}^{XZ(YZ)}(\bk)\, \hat{\tau}_a \otimes \hat{\sigma}_b,
$
where  the $\tilde{h}_{ab}^{XZ(YZ)}(\bk)$ are real even functions of momentum, and both $\hat{\tau}_a$ and $\hat{\sigma}_b$ are Pauli matrices encoding the orbital and the spin DOF, respectively ($\hat{\sigma}_0$ and $\hat{\tau}_0$ are identity matrices).
There are, in principle, 16 pairs of indexes $(a,b)$ but in  the presence of time-reversal and inversion symmetries these are constrained to only six. The symmetry-allowed terms are listed in Table \ref{Tab:H02orb}, accompanied by the explicit form of the dominant terms for each plane, as well as the associated physical process.
 
 \begin{table}[h]
 \caption{Symmetry-allowed $(a,b)$ indexes in the two-orbital normal state Hamiltonians along the $XZ$- and $YZ$-planes and explicit form of dominant terms for each plane. Here we follow the notation introduced in \cite{Suh:2020} and define $t'= \frac{2}{3}(t_x^{11}+t_y^{11}+t_x^{33})$, $t'_x = t_x^{11}-t_x^{33}$, $t'_y = t_y^{11}-t_x^{33}$, and $\tilde{t}= 8 t_{56z}^{SOC}$. The zero entries are defined by symmetry on the XZ and YZ planes.}
\begin{center}
\def\arraystretch{1.25}
    \begin{tabular}{cccc}
\br
     $(a,b)$ & $\tilde{h}_{ab}^{XZ}(\bk)$ & $\tilde{h}_{ab}^{YZ}(\bk)$ & Physical process  \\ \mr
     $(0,0)$  & $t' [1 + \cos (k_x)]$ & $t' [1 + \cos (k_y)]$ & intra-orbital hopping \\
     $(1,0)$  &  0 & 0 & inter-orbital hopping \\ 
     $(2,1)$  & $\eta$ & 0 & atomic SOC  \\ 
     $(2,2)$  & 0  & $\eta$ & atomic-SOC \\ 
     $(2,3)$  & $-\tilde{t} \sin(k_x/2)\sin(k_z/2) $  & $-\tilde{t} \sin(k_y/2)\sin(k_z/2) $ & $\bk$-SOC \\ 
     $(3,0)$  & $t'_y + t'_x \cos(k_x)$ & $t'_y + t'_x \cos(k_y)$ & intra-orbital hopping \\ \br
    \end{tabular}
\end{center}
\label{Tab:H02orb}
\end{table} 

\section{Local order parameters}

In analogy to the normal state Hamiltonian, we parametrize the superconducting order parameters for the three orbital model as
$
\hat{\Delta}(\bk) = \sum_{a,b}d_{ab}(\bk)\, \hat{\lambda}_a \otimes \hat{\sigma}_b\, (i\hat{\sigma}_2),
$
with momentum-dependent functions $d_{ab}(\bk)$ carrying information about its magnitude and phase. Each order parameter component is labelled by a pair of indexes $[a,b]$ (note the different brackets used for the parametrization of the terms in the normal state Hamiltonian and for the order parameter).  In the three orbital model, the proposed chiral d-wave state is dominated by \cite{Suh:2020}:
\begin{eqnarray}
\hat{\Delta}_{E_g}(\bk) = d_0 [\hat{\lambda}_5 \otimes \hat{\sigma}_3 + i \hat{\lambda}_6 \otimes \hat{\sigma}_3] (i\hat{\sigma}_2).
\end{eqnarray}
In the microscopic basis this order parameter is momentum independent (or s-wave) with magnitude $d_0$,  antisymmetric in the orbital DOF (given the explicit form of the Gell-Mann matrices $\lambda_5$ and $\lambda_6$), and a spin triplet state, with the d-vector along the z-axis.

Along the $XZ (YZ)$ plane we can write the local gap matrices as 
$
\hat{\Delta}^{XZ(YZ)} (\bk)= \sum_{ab}d_{ab}^{XZ(YZ)}(\bk)\, \hat{\tau}_a \otimes \hat{\sigma}_b\, (i\hat{\sigma}_2)
$. 
Each order parameter component is labelled by a pair of indexes with a label concerning the plane $[a,b]_p$, with $p=\{XZ,YZ\}$. Here we focus on local, s-wave, order parameters, which should satisfy the condition imposed by fermionic anti-symmetry, $
\hat{\Delta}(\bk) = - \hat{\Delta}^T(-\bk) \Rightarrow \hat{\Delta}= - \hat{\Delta}^T
$, 
 so we consider only order parameters associated with anti-symmetric matrices. These are listed in Tab. \ref{Tab:Fit}.

The orbitally antisymmetric order parameter, when projected into the XZ  (YZ) plane takes the form $[2,3]_{XZ(YZ)}$, as $\hat{\tau}_2$ is the only anti-symmetric 2$\times$2 matrix. More specifically, the $[6,3]$ component corresponding to an antisymmetric pair of $d_{xz}$ and $d_{xy}$ orbitals is associated with $[2,3]_{XZ}$ along the $XZ$ plane. Conversely, the $[5,3]$ component corresponding to an antisymmetric pair of $d_{yz}$ and $d_{xy}$ orbitals is associated with $[2,3]_{YZ}$ along the $YZ$ plane. 

\section{Origin of line nodes}

For this discussion, the simple 2$\times$2 models along the planes $p=\{XZ,YZ\}$ are going to be useful so we can proceed analytically. As the matrix structures are the same for both planes, we drop the $p$ superscript from here on. In order to determine the gap in the band basis, it is useful to look directly at the eigenvalaues of the BdG Hamiltonian:
\begin{eqnarray}
\hat{H}_{BdG}(\bk)=\begin{pmatrix}
\hat{H}_0(\bk) & \hat{\Delta}(\bk)\\
\hat{\Delta}^\dagger(\bk)&- \hat{H}_0^*(-\bk)
\end{pmatrix},
\end{eqnarray}
where $\hat{H}_0(\bk)$ is the normal-state Hamiltonian and $\hat{\Delta}(\bk)= \Delta_0 \hat{\tau}_c \otimes \hat{\sigma}_d (i \hat{\sigma}_2)$, with different gaps associated with distinct subscripts $[c,d]$, as listed in Table \ref{Tab:Fit}. Here we already assume the gap to be momentum-independent with magnitude $\Delta_0$. The eigenvalues can be explicitly written as
\begin{eqnarray}
E^{\pm,\pm}(\bk) = \pm \sqrt{\tilde{h}_{00}^2(\bk) + \tilde{\bh}^2(\bk)+ \Delta_0^2 \pm 2 \sqrt{\tilde{h}_{00}^2(\bk)\tilde{\bh}^2(\bk) + \Delta_0^2 f^2_C(\bk) }},
\end{eqnarray}
where $\tilde{\bh}(\bk) = (\tilde{h}_{10}(\bk),\tilde{h}_{30}(\bk),\tilde{h}_{21}(\bk),\tilde{h}_{22}(\bk),\tilde{h}_{23}(\bk))$ is a five-dimensional vector. The function
$
f^2_C(\bk)= \frac{1}{16} \text{Tr} [\tilde{F}_C^{0\dagger} (\bk) \tilde{F}_C^0(\bk)],
$
is written in terms of the normalized superconducting fitness function $\tilde{F}_C^0(\bk) = \hat{F}_C^0(\bk)/\Delta_0$, with
$
\hat{F}_C^0(\bk) = \hat{H}_0(\bk) \hat{\Delta}(\bk) - \hat{\Delta}(\bk) \hat{H}_0^*(-\bk)
$. Note that the normal-state eigenenergies are given by
$
E_0^{\pm} (\bk)= \tilde{h}_{00}(\bk) \pm |\tilde{\bh}(\bk)|
$,
such that the Fermi surface is defined by $\tilde{h}_{00}^2(\bk)  = \tilde{\bh}^2(\bk)$. Note also that if $f^2_C(\bk) =0$, the eigenenergies read $E^{\pm,\pm}(\bk) = \pm \sqrt{(|\tilde{h}_{00}(\bk)| \pm |\tilde{\bh}(\bk)|)^2+ \Delta_0^2}$, and the superconducting state is fully gapped.

In case $f^2_C(\bk) \neq 0$, the energy spectra in the superconducting state at the momenta belonging to the normal-state Fermi surface can then be written as
\begin{eqnarray}
E^{\pm,\pm}(\bk)\Big|_{FS} = \pm \sqrt{2 \tilde{h}_{00}^2(\bk) + \Delta_0^2 \pm 2 \sqrt{\tilde{h}_{00}^4(\bk) + \Delta_0^2 f^2_C(\bk) }}.
\end{eqnarray}
If we now expand for small $\Delta_0$, we find for the low lying bands:
\begin{eqnarray}
E^{\pm,-}(\bk)\Big|_{FS} \approx \pm \Delta_0  \sqrt{ \left(1- \frac{f_C^2(\bk)}{\tilde{h}_{00}^2(\bk)}\right)},
\end{eqnarray}
and we can
define the gap structure in the band basis as:
\begin{eqnarray}\label{Eq:GapB}
\mathcal{D}_B(\bk) =\left[ E^{+,-}(\bk)\Big|_{FS} - E^{-,-}(\bk)\Big|_{FS}\right]/2 \approx  \frac{\Delta_0}{|\tilde{h}_{00}(\bk)|}  \sqrt{\tilde{h}_{00}^2(\bk)- f_C^2(\bk)}.
\end{eqnarray}

This definition is in direct analogy to the single band problem with normal state dispersion $\epsilon(\bk)$. The dispersion in the superconducting state is $E^\pm (\bk) = \pm \sqrt{\epsilon^2(\bk) + \Delta^2}$, so we can identify the gap at the Fermi surface by taking $\epsilon(\bk) =0$, and defining $[E^+ (\bk) |_{FS}-E^- (\bk) |_{FS}]/2 = |\Delta|$. 
 
From Eq. \ref{Eq:GapB}, it is clear that the condition for the presence of nodes at the normal state Fermi surface is:
\begin{eqnarray}
\tilde{h}_{00}^2(\bk) - f_C^2(\bk)=0 \hspace{1cm} \text{or} \hspace{1cm} \tilde{\bh}^2(\bk) - f_C^2(\bk)=0.
\end{eqnarray}
These equations indicate that there can be nodes in the superconducting spectrum of Sr$_2$RuO$_4$ given the nontrivial form factors originated from the normal-state Hamiltonian $\hat{H}_0(\bk)$, even though the gap in the orbital basis is isotropic (s-wave). Note that the fitness function $\hat{F}^0_C(\bk)$ appears in this context as a key parameter defining the nodes, reinforcing its importance for the understanding of the phenomenology of superconducting states in complex materials \cite{Ramires:2016,Ramires:2017,Ramires:2018,Ramires:2019,Andersen:2020, Suh:2020}. As a final note, keeping higher order terms in $ \Delta_0^2$ in the expansion above, we find structures associated with Bogoliubov Fermi surfaces, which can be understood as inflated nodes \cite{Agterberg:2017, Brydon:2018}. We are not going to discuss these further here.

In Tab. \ref{Tab:Fit} we summarise the fitness functions for each possible local order parameter labelled by $[a,b]$ in the two-orbital models (valid for both XZ and YZ planes). Based on the explicit form of the $\tilde{h}_{ab}(\bk)$ functions given in Tab. \ref{Tab:H02orb}, we can directly infer the presence and position of nodes. We highlight the presence of horizontal line nodes for the order parameter labelled as $[2,3]_p$, corresponding to the two components of the chiral d-wave order parameter $\{[5,3],[6,3]\}$  in the three orbital model for $p=\{YZ,XZ\}$, respectively. It can be shown that the nodes are not confined only to the $XZ$ and $YZ$ planes, but appear for any point with $k_z=\{0, \pm 2\pi \}$. Note that both components present horizontal line nodes, so even a complex superposition of the two components carries a horizontal line node. This is directly related to the fact that both $\tilde{h}_{23}^{XZ(YZ)}$ are proportional to $\sin(k_z/2)$.

\begin{table}[h]
        \caption{\label{Tab:Fit} Normal state parameters $\tilde{h}_{ab}(\bk)$ determining the nodal structure for the s-wave gaps in the two-orbital effective model. These results are valid for both $XZ$ and $YZ$ planes, therefore we drop here the superscript $p$. The nodes as discussed assuming cylindrical Fermi surfaces. We neglect the possibility of fine-tuned accidental nodes that can appear for specific positions of the Fermi surface. Here we omit the explicit momentum dependence for conciseness, and used the identity valid at the Fermi surface $\tilde{h}_{00}^2= \tilde{\bh}^2$.}
        \begin{center}
        \centering
\begin{tabular}{cccc}
\br
 Gap Matrix &  $f_C^2$ & $|\tilde{\bh}^2 - f_C^2|$ & Line Nodes \\
 \mr
$[0,0]$ & 0 & $\tilde{h}_{00}^2$ & No \\
$[1,0]$ & $\tilde{h}_{21}^2+\tilde{h}_{22}^2+\tilde{h}_{23}^2+\tilde{h}_{30}^2$ & $\tilde{h}_{10}^2$ & Vertical \\
$[2,1]$ & $\tilde{h}_{10}^2+\tilde{h}_{22}^2+\tilde{h}_{23}^2+\tilde{h}_{30}^2$ & $\tilde{h}_{21}^2$ & No (XZ)/Vertical (YZ) \\
$[2,2]$ & $\tilde{h}_{10}^2+\tilde{h}_{21}^2+\tilde{h}_{23}^2+\tilde{h}_{30}^2$ & $\tilde{h}_{22}^2$ & Vertical (XZ)/No (YZ) \\ 
$[2,3]$ & $\tilde{h}_{10}^2+\tilde{h}_{21}^2+\tilde{h}_{22}^2+\tilde{h}_{30}^2$ & $\tilde{h}_{23}^2$ & Horizontal \\ 
$[3,0]$ & $\tilde{h}_{10}^2+\tilde{h}_{21}^2+\tilde{h}_{22}^2+\tilde{h}_{23}^2$ & $\tilde{h}_{30}^2$ & No \\ \br
    \end{tabular}
    \end{center}
\end{table}

\section{Origin of gap minima along the diagonals}

\begin{center}
\begin{figure}[h]
\begin{center}
\includegraphics[width=\linewidth, keepaspectratio]{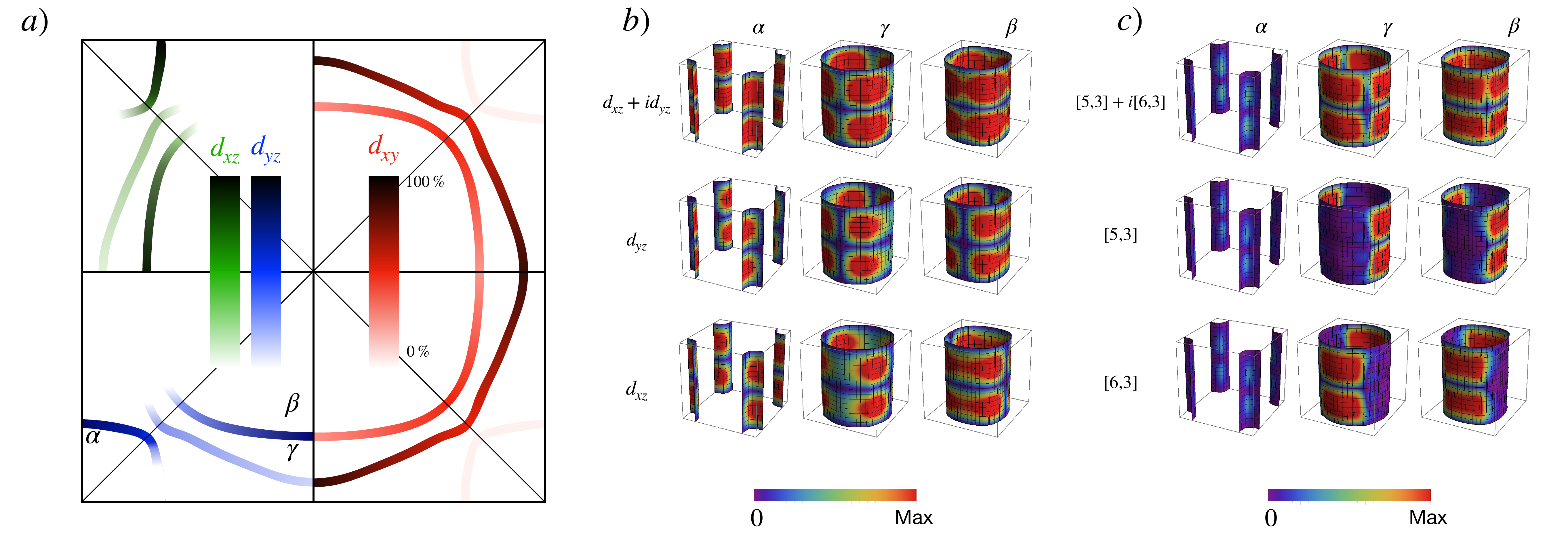}
\end{center}
\caption{a) Schematic orbital content over the Fermi surfaces (FSs) of Sr$_2$RuO$_4$ in the $k_z=0$ plane. The color (red, blue, green) encode the orbital ($d_{xy}$, $d_{yz}$, $d_{xz}$) content as low (bright) or high (dark) in each Fermi surface sheet (labelled as $\alpha$, $\beta$ and $\gamma$). Adapted from \cite{Veenstra:2014,Tamai:2019}. b) Top: Naive chiral d-wave state $\propto (\sin k_x+ i\sin k_y)\sin k_z  $ superimposed on the three-dimensional FSs of Sr$_2$RuO$_4$. Note the horizontal line nodes and the minima along the x- and y-directions. Middle (Bottom): separate plot of the component $\sin k_x \sin k_z$ ($\sin k_x \sin k_z$). Note the nodes along the x-(y-) direction. c) Top: Inter-orbital s-wave chiral superconducting state gap (encoded in the matrix structure $[5,3]+i[6,3]$) on the three-dimensional FSs of Sr$_2$RuO$_4$. Note the horizontal line nodes and the minima along the diagonals. Middle (Bottom): separate plot of the component $[5,3]$ ($[6,3]$). Note the extended regions with very small gap associated with the absence of certain orbitals on these regions of the Fermi surface.}
\label{Fig:Gaps}
\end{figure}
\end{center}

A second feature of the chiral d-wave order parameter dominated by the $\{[5,3],[6,3]\}$ components is the presence of gap minima along the diagonals, reminiscent of a $d_{x^2-y^2}$ structure. This can be understood by the fact that the two-component aspect of the order parameter in this case is not carried by the basis functions $\sin(k_x)\sin(k_z)$ and $\sin(k_y)\sin(k_z)$, but is carried by the orbital content of the gap structure. The orbital content is encoded by the Gell-Mann matrices. These matrices encode specific types of pairing: $\lambda_5$ corresponds to antisymmetric pairing between $d_{yz}$ and $d_{xy}$ orbitals, while $\lambda_6$ corresponds to antisymmetric pairing between $d_{xz}$ and $d_{xy}$ orbitals. In this case, the orbital content on the Fermi surfaces is going to limit the magnitude of the gap along specific directions. In Fig. \ref{Fig:Gaps} we compare the gap structure expected for a $d_{xz}+id_{yz}$ order parameter (accompanied by the $[0,0]$ matrix structure), displaying gap maxima along the diagonals, with the gap structure associated with the $[5,3]+i [6,3]$ structure, with gap minima along the diagonals.

\ack The author thanks Yoshi Maeno for the question that inspired this proceeding. The author also thanks Han Gyeol Suh, Henri Menke, Philip M. R. Brydon, Carsten Timm, and Daniel F. Agterberg for discussions and previous collaborations that motivated this work. The author acknowledges the financial support of the Swiss National Science Foundation through an Ambizione Grant.

\section*{References}
\bibliographystyle{unsrt}
\bibliography{Gap}{}



\end{document}